\documentclass[a4paper]{jpconf}
\usepackage{graphicx}
\usepackage[center]{subfigure}
\usepackage{xspace}
\usepackage{footnote}
\usepackage{amsmath}
\usepackage{cite}   

\newcommand{\camb}{\texttt{camb}\xspace}
\newcommand{\cosmomc}{\texttt{CosmoMC}\xspace}
\newcommand{\hub}{\ensuremath{\mathcal{H}}}
\newcommand{\lcdm}{$\Lambda$CDM}
\newcommand{\wla}{\ensuremath{w_\Lambda}}

\newcommand{\rhodm}{\ensuremath{\rho_{\rm{DM}}}}
\newcommand{\rhode}{\ensuremath{\rho_\Lambda}}

\newcommand{\logA}{\ensuremath{\log(10^{10}A_s)}}

\newcommand{\Hou}{\ensuremath{\, \text{Km s}^{-1}\text{ Mpc}^{-1}}}

\newlength{\lenerrbars}
\newlength{\lencontours}
\newlength{\lencls}
\begin{document}
\title{Constraining the interaction between dark matter and dark energy with CMB data}

\author{Riccardo Murgia}

\address{SISSA, via Bonomea 265, 34136 Trieste, Italy}
\address{Department of Physics, University of Torino, Via P. Giuria 1, I--10125 Torino, Italy}
\address{INFN, Sezione di Torino, Via P. Giuria 1, I--10125 Torino, Italy}

\ead{riccardo.murgia@sissa.it}

\begin{abstract}
We briefly discuss the intriguing case of a phenomenological non-gravitational coupling in
the dark sector, where the interaction is parameterized as an energy transfer either
from dark matter to dark energy or the opposite.
We show that a non-zero coupling with an energy flow from the latter to the former
leads to a full reconciliation of the tension between high- and low-redshift observations
present in the standard cosmological model.
\end{abstract}

\section{Introduction}
\label{sec:intro}
\looseness=-1
The cosmic microwave background (CMB) radiation ani\-so\-tro\-pies represent 
one of the pillars of modern cosmology. They provide a convincing picture that the Universe is 
dominated by ``dark'' components, whose presence is felt only through their gravitational effects. The results 
of the Planck Collaboration~\cite{Adam:2015rua} show that baryonic matter accounts only for 5\% of the total energy density today,
while 26\% is accounted for by Dark Matter (DM) and 69\% is in the form of a diffuse component, 
the Dark Energy (DE), responsible of the accelerated expansion of the Universe. 
Furthermore, a whole set of independent cosmological observations 
(such as high-redshift supernovae, Baryon Acoustic Oscillations (BAOs), Redshift Space Distortions (RSDs) and gravitational lensing) contribute to what 
has emerged as a successful interpretation of the behavior of our Universe, namely the $\Lambda$CDM model. Some tension exists, though, on the cosmological 
and local determination of the Hubble constant $H_0$ and on the amplitude of the linear power spectrum on the scale of 8 $h^{-1}$ Mpc, $\sigma_8$, 
as discussed for example in~\cite{Ade:2015xua}. 

\looseness=-1
The nature of DM and DE is still unknown and while DM can be naturally interpreted as being formed by a new type of elementary particle, 
a clear view on the solution of DE is still lacking. The simplest interpretation of DE is in the form of a cosmological constant $\Lambda$. 
In standard cosmology, based on general relativity, the cosmological constant can be seen as a fluid endowed 
with equation of state (EoS) $p_\Lambda=\wla\rhode$, 
with $\wla=-1$ ($p_\Lambda$ and $\rhode$ being the pressure and the energy density of DE, respectively): 
this allows to explain the current accelerated phase of the Universe.
\looseness=-1
The most economic approach is that the two sectors, i.e.~DM and DE, 
do not interact other than gravitationally. However, they might well have some form of non-gravitational coupling, 
with relevant implications both at the fundamental level and at the cosmological level, 
since cosmological observables can be affected by the presence of an interaction between DM and DE (see e.g.~\cite{Wang:2016lxa}).
At the same time, cosmological and astrophysical observations can be proficiently 
used to constrain any form of coupling between the two dark components.

\section{Parameterization}
\label{sec:param}
This new interaction can be phenomenologically introduced in various ways (see e.g.~\cite{Koyama:2009gd} for a classification). 
In our analysis~\cite{Murgia:2016ccp}, which follows similar approaches
\cite{Abdalla:2014cla,Costa:2013sva,Gavela:2010tm,Salvatelli:2013wra}, 
we phenomenologically parameterize the coupling between DM and DE through an energy transfer from one sector to the other. 
This can be expressed through the non-conservation of their stress-energy tensors $T^{\mu\nu}_i$ ($i=$ DM, DE). 
In this approach, the stress-energy tensors of DM and DE are not separately conserved (while the total stress-energy tensor of DM and DE is), 
and we parameterize this as:
\begin{subequations}
\begin{eqnarray}
 \nabla_\mu T^{\mu\nu}_{\rm{DM}} & = & +Q u^\nu_{\rm{DM}}/a\label{eq:stressenergyTDM}\,,\\
 \nabla_\mu T^{\mu\nu}_{\rm{DE}} & = & -Q u^\nu_{\rm{DM}}/a\label{eq:stressenergyTDE}\,,
\end{eqnarray}
\end{subequations}
where the coefficient $Q$ encodes the interaction between the two sectors, 
$u^\nu_{\rm{DM}}$ is the DM four-velocity and 
$a$ is the time-dependent scale factor of the Universe
\cite{Abdalla:2014cla,Costa:2013sva,Gavela:2010tm,Salvatelli:2013wra}. 
The ensuing evolution equations for the DM and DE energy densities are therefore:
\begin{subequations}\label{eq:dmdecou}
  \begin{eqnarray}
 \dot{\rho}_{\rm{DM}}+3\hub\rhodm &=& +Q \;, \label{eq:dmdecou_dm}\\
 \dot{\rhode}+3\hub(1+\wla)\rhode &=& -Q \,.  \label{eq:dmdecou_de}
  \end{eqnarray}
\end{subequations}
If $Q>0$ the energy transfer is from DE to DM, i.e.~DE decays into DM, whereas 
if $Q<0$ the energy flux has the opposite direction, i.e.~DM decays into DE.
In this work, we will not focus on specific theoretical frameworks, 
we instead use a phenomenological approach and we study one specific model, 
where the coupling term is proportional to the DE density,
\begin{equation}
\label{eq:coupl}
Q=\xi \hub \rhode\,,
\end{equation}
\looseness=-1
where $\xi$ is a dimensionless coupling parameter and
the time dependence of the interaction rate is governed by the Hubble rate 
$\hub = \dot{a}/a$~\cite{Costa:2013sva,Salvatelli:2013wra}. For $\xi=0$ we recover the uncoupled case
of standard cosmology.
\looseness=-1
As we will show below, this model offers the possibility to loosen the tension between the local and cosmological determinations 
of the Hubble constant $H_0$ and the matter fluctuations parameter $\sigma_8$.
\looseness=-1
Our baseline model is the \lcdm~model, where $\Lambda$ refers to the DE and CDM stands for
Cold Dark Matter.
\looseness=-1
It can be described using six free parameters:
the present baryon density $\Omega_bh^2$;
the present CDM density $\Omega_ch^2$;
the ratio of the sound horizon to the angular diameter distance at decoupling $\theta$;
the optical depth at reionization $\tau$;
the amplitude $A_s$ and
the spectral index $n_s$ 
of the primordial power spectrum of scalar perturbations, at the pivot scale $k=0.05$ Mpc$^{-1}$.
The DE density is a derived parameter, under the assumption of a flat Universe.
The current value of the Hubble parameter $H_0$ and
the root-mean-square fluctuations in total matter in a sphere of $8h^{-1}$~Mpc radius,
$\sigma_8$, are derived parameters too.
We adopt flat priors in the ranges listed in Table~\ref{tab:priorslcdm}.
  
\begin{savenotes}
\begin{table}[b]
\begin{center}\small{
\begin{tabular}{|c|c|c|}
  \hline
  Parameter		& Prior		\\	\hline	
  $\Omega_bh^2$ 	& [0.005, 0.1]	\\
  $\Omega_ch^2$ 	& [0.001, 0.5]	\\
  $100\theta$ 		& [0.5, 10]	\\ 	
  $\tau$  		& [0.01, 0.8]	\\ 
  $\logA$ 		& [2.7, 4]	\\ 	
  $n_s$ 		& [0.9, 1.1]	\\ 	\hline
  $\sum m_{\nu}$	& 0.06 eV	\\
  $N_{\nu}$		& 3.046		\\ 	\hline
  $H_0$~[\Hou]		& [20,100]	\\	\hline
\end{tabular}}
\caption{Priors and constraints on the parameters adopted in the analysis. All priors are flat in the listed intervals.}
\label{tab:priorslcdm}
\end{center}
\end{table}
\end{savenotes}
\looseness=-1
The evolution of the DM and DE densities can be obtained by solving
Eqns.~\eqref{eq:dmdecou_dm} and \eqref{eq:dmdecou_de}:
\begin{subequations}
\label{eq:coupl_bg}{
  \begin{eqnarray}
  \rhodm &=& \rho_{\rm{DM}}^0\,a^{-3} + {\rhode^0 a^{-3} \Bigg[\frac{\xi}{3\wla+\xi}
  \big(1-a^{-3\wla-\xi}\big)\Bigg]}\label{eq:coupl_bgDM}\,,\\
  \rhode &=& \rhode^0\, a^{-3(\wla+1)-\xi}\,,\label{eq:coupl_bgDE}
  \end{eqnarray}}
\end{subequations}
%
where $\rho_{i}^0$ ($i=$ DM, DE) is the energy density of the species $i$ today.
We recall that $\xi<0$ corresponds to an energy flux from DM to DE, i.e.~DM decaying into DE, 
while $\xi>0$ corresponds to an energy flux from DE to DM, i.e.~DE decaying into DM.
Hereafter, we will refer to the former case as Model 1 (MOD1) and to the latter case as Model 2 (MOD2).

\looseness=-1
Looking at Eqns.~\eqref{eq:coupl_bgDM} and \eqref{eq:coupl_bgDE}, we notice that 
it is difficult to disentangle the effects of the DE EoS parameter $\wla$
from the coupling $\xi$ by only studying the background evolution; 
we must include the perturbation evolution equations, which are also affected by the DM/DE coupling.
The perturbation equations in the linear regime can be expressed in the synchronous gauge as~\cite{Costa:2013sva}:
\begin{subequations}
\label{eq:pertcou}{
\begin{eqnarray}
\dot{\delta}_{\rm{DM}}
& = 
&-\left(kv_{\rm{DM}}+\frac{\dot{h}}{2}\right)+ {\xi \hub \frac{\rhode}{\rhodm}(\delta_\Lambda-\delta_{\rm{DM}})}
\,;\label{eq:pertcouDMd}\\ 
\dot{v}_{\rm{DM}}& = &-\hub v_{\rm{DM}}\left(1+{\xi\frac{\rhode}{\rhodm}}\right)
\,;\label{eq:pertcouDMv}\\ 
\dot{\delta}_{\Lambda}
& = 
&-(1+\wla)\left(kv_\Lambda+\frac{\dot{h}}{2}\right)
-3\hub(1-\wla)\cdot \left(\delta_\Lambda \hub(3(1+\wla)+{\xi})\frac{v_\Lambda}{k}\right)
\,;\label{eq:pertcouDEd}\\ 
\dot{v}_{\Lambda}& = &-2\hub\left(1+{\frac{\xi}{1+\wla}}\right)v_\Lambda+k\frac{\delta_\Lambda}{1+\wla}
\,;\label{eq:pertcouDEv}
\end{eqnarray}}
\end{subequations}
where $h=6\phi$ is the synchronous gauge metric perturbation 
and ${v}_{\rm{DM}}$,~i.e. the DM peculiar velocity, is fixed to zero using the gauge freedom.
Moreover, the DE sound speed is fixed: $c_{s,\Lambda}=1$.

\looseness=-1
For our analysis, we implemented the modified equations 
into the numerical Boltzmann solver \camb~\cite{Lewis:1999bs}
and we modified the Markov Chain Monte Carlo (MCMC) code \cosmomc~\cite{Lewis:2002ah} 
in order to include $\xi$ as an additional parameter.
The parameters specific to the DM/DE coupling, $\xi$ and $\wla$, are variable inside the ranges listed in Table~\ref{tab:priorscde};
we assumed flat priors.

\begin{table}[t]
\begin{center} \small{
\begin{tabular}{|c|c|c|c|}
  \hline
  	& \multicolumn{3}{c|}{Prior}		\\ \hline
  Parameter	& \lcdm	& MOD1		& MOD2 	\\ \hline
  $\wla$  	&$-1$	& $[-0.999, -0.1]$& $[-2.5, -1.001]$\\
  $\xi$ 	& $0$	& $[-1, 0]$	& $[0, 0.5]$	\\ \hline
  &no interaction& DM decays into DE & DE decays into DM \\ \hline
\end{tabular}}
\caption{Priors on 
the coupling parameter $\xi$ and the DE equation of state parameter $\wla$. Both are flat in the listed intervals,
which are chosen in such a way to avoid early time instabilities and unphysical values for $\rho_{\rm{DM}}$ (see~\cite{Murgia:2016ccp} for details).}
\label{tab:priorscde}
\end{center}
\end{table}

\section{Constraints from cosmological data}
\label{sec:constr}
We base our analysis on CMB data from the latest Planck release~\cite{Adam:2015rua}. 
Specifically, we consider as our minimal data combination
the full temperature autocorrelation spectrum in the range $2\leq\ell\leq2500$ (denoted as ``PlanckTT'') plus the
low-$\ell$ Planck polarization spectra in the range $2 \leq\ell\leq 29$
(denoted as ``lowP'')~\cite{Aghanim:2015xee}.
Additionally, we consider and add separately 
the high-$\ell$ Planck polarization spectra in the range
$30\leq\ell< 2500$ (hereafter ``highP'')~\cite{Aghanim:2015xee}.

\looseness=-1
Since the coupling between DE and DM introduces a time dependence in the background evolution of DE and DM 
(see Eqns.~\eqref{eq:coupl_bgDM} and \eqref{eq:coupl_bgDE}), 
it is important to test our models using data at different redshifts with respect to CMB measurements.
\looseness=-1
Thus, we consider the luminosity distances of SNIa 
from the SNLS and SDSS catalogs as re-analyzed in~\cite{Betoule:2014frx} (``JLA'' hereafter).
We also include the BAO as determined by
6dFGS~\cite{Beutler:2011hx}, 
SDSS-MGS~\cite{Ross:2014qpa} and
BOSS DR11~\cite{Anderson:2013zyy},
together with RSD data from~\cite{Samushia:2013yga}.
\looseness=-1
We will refer to the combination of these measurements as to the ``BAO/RSD'' dataset.
\looseness=-1
We also include information on the power spectrum of the lensing potential reconstructed by Planck 
from the trispectrum measurement~\cite{Ade:2015zua} (hereafter ``lens'').

It will be interesting to compare our results with the weak lensing determinations obtained from the cosmic shear measurements 
of the CFHTLenS survey~\cite{Heymans:2012gg}, that appear to be in substantial tension with the Planck results~\cite{Raveri:2015maa}.
The tension can be explained invoking the presence
of some unaccounted systematics in the analyses of the experimental data
or an incomplete modeling of the theoretical predictions,
but can also be the result of the existence of new physics beyond the \lcdm~model.
In this respect, it is worth to discuss also other experiments that probe the mass distribution at late times, 
such as cluster counts through the Sunyaev-Zel'dovich (SZ) effect from Planck~\cite{Ade:2015fva}.
As mentioned before, a clear detection of a preference for a low $\sigma_8$ from the local determinations with respect to the CMB results
would indicate the possible existence of new physics beyond the \lcdm~model.

We do not include in our analyses constraints on the Hubble parameter $H_0$, the expansion rate of the Universe today,
due to the tensions that exist between local determinations and CMB estimates for this observable.
Planck constraints in the context of the \lcdm~model are typically lower than the local measurements. 
Therefore we confront and discuss our results in comparison with the local determination of $H_0$~\cite{Efstathiou:2013via,Riess:2016jrr}.
The significance of the tension depends on the distance calibration. Only the result obtained in~\cite{Efstathiou:2013via}
is consistent with the CMB result within 1$\sigma$, while the other measurements present some tension.

We will show that MOD2 can resolve the tension between local and cosmological determinations of both $H_0$ and $\sigma_8$.

\looseness=-1
In our analyses we have explored different combinations of the listed datasets: 
our starting point is the CMB-only dataset ``PlanckTT+lowP'', 
while we will indicate with ``ALL'' a combination involving all the mentioned datasets.
For each data combination we will test the three cosmological models (\lcdm, MOD1, MOD2) to investigate
the impact of the coupled scenarios.

\begin{table}[t]                                                                           
\begin{center} \small{                                                                         
\begin{tabular}{|c|c|c|c|}                                                              
\hline
Parameter &	\lcdm	& MOD1	& MOD2\\
\hline
$100\Omega_bh^2$& $2.222\,^{+0.047}_{-0.043}$    & $2.216\,^{+0.046}_{-0.045}$    & $2.226\,^{+0.047}_{-0.046}$    \\
$\Omega_ch^2$	& $0.120\,^{+0.004}_{-0.004}$    & $0.069\,^{+0.051}_{-0.062}$    & $0.133\,^{+0.018}_{-0.015}$    \\
$100\theta$	& $1.0409\,^{+0.0009}_{-0.0009}$ & $1.0441\,^{+0.0051}_{-0.0037}$ & $1.0402\,^{+0.0013}_{-0.0013}$ \\
$\tau$		& $0.078\,^{+0.039}_{-0.037}$    & $0.077\,^{+0.039}_{-0.038}$    & $0.077\,^{+0.039}_{-0.038}$    \\
$n_s$		& $0.965\,^{+0.012}_{-0.012}$    & $0.964\,^{+0.013}_{-0.012}$    & $0.966\,^{+0.013}_{-0.012}$    \\
$\logA$		& $3.089\,^{+0.074}_{-0.072}$    & $3.088\,^{+0.073}_{-0.073}$    & $3.087\,^{+0.073}_{-0.074}$    \\ 	\hline
$\xi$		& $0$                            & $(-0.790,0]$                   & $[0,0.269)$                    \\
$\wla$		& $-1$                           & $[-1,-0.704)$                  & $-1.543\,^{+0.515}_{-0.436}$   \\ 	\hline
$H_0$~[\Hou]	& $67.28\,^{+1.92}_{-1.89}$      & $67.91\,^{+7.26}_{-7.54}$      & $>68.31$                       \\
$\sigma_8$	& $0.830\,^{+0.029}_{-0.028}$    & $1.464\,^{+1.917}_{-0.834}$    & $0.898\,^{+0.163}_{-0.160}$   
\\ 	\hline
\end{tabular}}
\caption{Marginalized limits at the 2$\sigma$ C.L.\ for the relevant parameters of this analysis. The results are
obtained using the ``PlanckTT+lowP'' dataset, for the three different models (\lcdm, MOD1 and MOD2).
When an interval denoted with parenthesis is given, it refers to the  2$\sigma$ C.L.\ range starting from the
prior extreme, listed in Tables~\ref{tab:priorslcdm} and~\ref{tab:priorscde}.}
\label{tab:cmb}
\end{center}
\end{table}

\begin{table}[t]                                                                           
\begin{center}                                                                          
\small{\begin{tabular}{|c|c|c|c|}                                                              
\hline
Parameter &	\lcdm	& MOD1	& MOD2\\
\hline
$100\Omega_bh^2$& $2.229\,^{+0.028}_{-0.028}$    & $2.228\,^{+0.030}_{-0.030}$    & $2.227\,^{+0.031}_{-0.030}$    \\
$\Omega_ch^2$	& $0.119\,^{+0.002}_{-0.002}$    & $0.091\,^{+0.028}_{-0.031}$    & $0.135\,^{+0.014}_{-0.014}$    \\
$100\theta$	& $1.0409\,^{+0.0006}_{-0.0006}$ & $1.0426\,^{+0.0021}_{-0.0018}$ & $1.0400\,^{+0.0010}_{-0.0010}$ \\
$\tau$		& $0.062\,^{+0.025}_{-0.025}$    & $0.063\,^{+0.027}_{-0.026}$    & $0.059\,^{+0.028}_{-0.027}$    \\
$n_s$		& $0.966\,^{+0.008}_{-0.008}$    & $0.966\,^{+0.009}_{-0.009}$    & $0.966\,^{+0.009}_{-0.009}$    \\
$\logA$		& $3.055\,^{+0.045}_{-0.046}$    & $3.058\,^{+0.049}_{-0.049}$    & $3.050\,^{+0.050}_{-0.051}$    \\ 	\hline
$\xi$		& $0$                            & $(-0.463,0]$                   & $0.159\,^{+0.146}_{-0.154}$    \\                                                   
$\wla$		& $-1$                           & $[-1,-0.829)$                  & $(-1.129,-1]$                  \\ 	\hline
$H_0$~[\Hou]	& $67.72\,^{+1.01}_{-0.97}$      & $67.57\,^{+1.81}_{-1.79}$      & $67.83\,^{+1.90}_{-1.75}$      \\
$\sigma_8$	& $0.812\,^{+0.017}_{-0.017}$    & $0.994\,^{+0.283}_{-0.202}$    & $0.749\,^{+0.067}_{-0.061}$    
\\ 	\hline
\end{tabular}}
\caption{The same as in Tab.~\ref{tab:cmb} but for the analysis on
the ``ALL'' dataset.
}
\label{tab:all}
\end{center}
\end{table}

\looseness=-1
The results of the analyses for the ``CMB only" and ``ALL" datasets are reported in Tables~\ref{tab:cmb} and~\ref{tab:all}, respectively.
The Tables show the 2$\sigma$ constraints for the relevant cosmological parameters. We find that most of them 
are not sensitive to the coupling in the dark sector and the ensuing results are quite unchanged when moving from \lcdm\ to MOD1 or MOD2. 
As expected, there is however a strong correlation between the coupling parameter $\xi$ and the current DM density $\Omega_ch^2$.
For $\xi<0$ (MOD1), the bigger is the interaction, the smaller is the DM abundance today,
i.e.~more DM decayed into DE during the evolution.
Conversely, for $\xi>0$ (MOD2) a larger DM density is predicted. This is manifest in Tables~\ref{tab:cmb} and~\ref{tab:all}.
Given a flat Universe, this turns out in different values for the DE energy density parameter 
today $\Omega_\Lambda$ in the different models.

\begin{figure}[b]
  \centering
  \includegraphics[page=1,width=\lencontours]{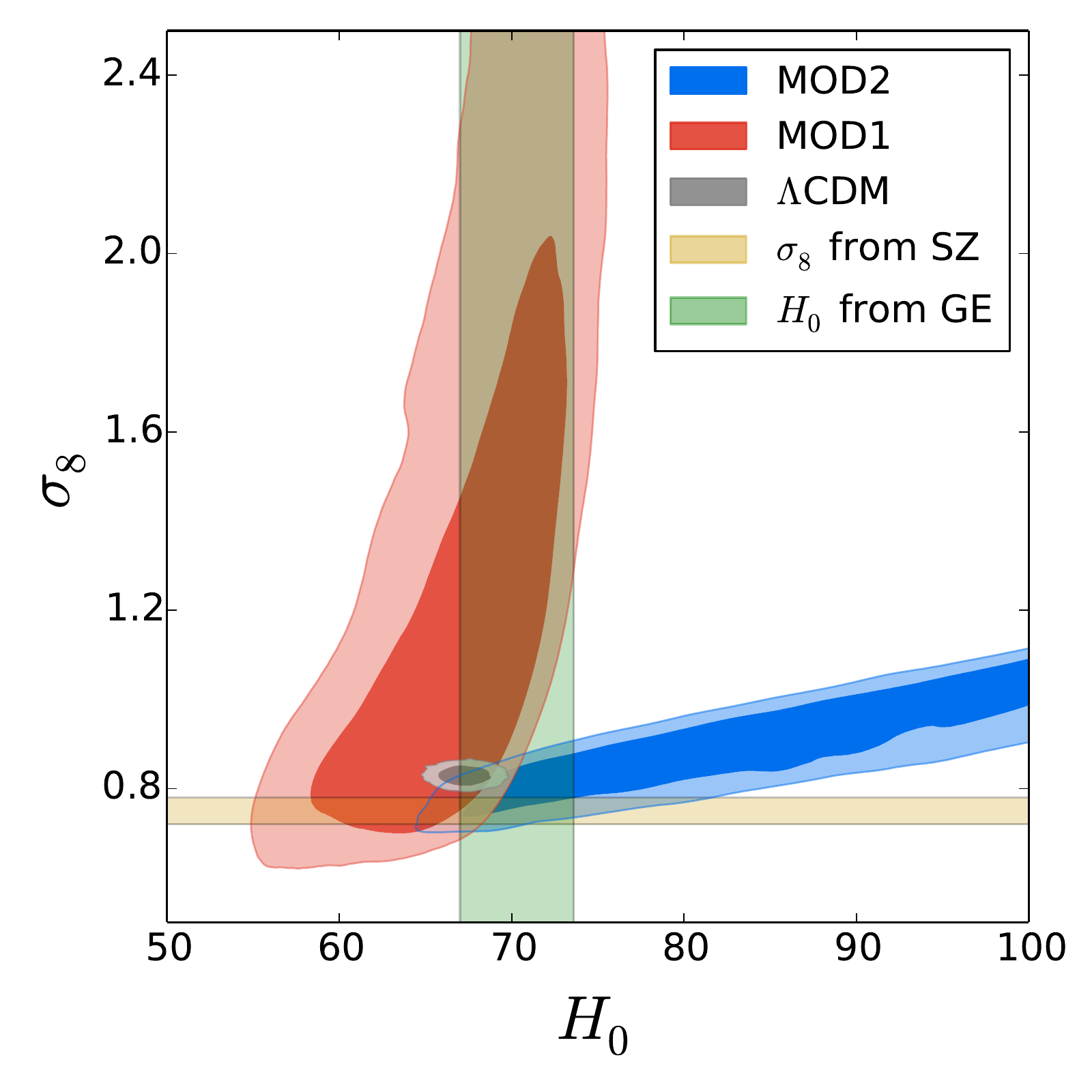}
  \includegraphics[page=2,width=\lencontours]{images/sigma8_h0.pdf}
  \caption{
  Marginalized 1$\sigma$ and 2$\sigma$ C.L.\ allowed regions in the ($\sigma_8$, $H_0$) plane for different models:
  \lcdm~(gray), MOD1 (red) and MOD2 (blue).
  The left panel refers to the CMB only dataset ``PlanckTT+lowP'', 
  while the right panel refers to the full combination considered here (``ALL'').
  The vertical green band denotes  the interval $H_0=70.6\pm3.3\Hou$ (GE)~\cite{Efstathiou:2013via}, 
  while the horizontal dark yellow band stands for $\sigma_8=0.75\pm0.03$ (SZ)~\cite{Ade:2013lmv}.
  }
  \label{fig:sigma8_h0}
\end{figure}

\looseness=-1
It is interesting to note that MOD1 predicts a value for $\sigma_8$ significantly larger than the \lcdm~prediction
(see both Tab.~\ref{tab:cmb} and Tab.~\ref{tab:all}).
Since MOD1 predicts a larger amount of DM in the early Universe, there is more clustering in the primordial Universe, 
that results in an earlier transition to the non-linear evolution and hence to an unavoidably larger value for $\sigma_8$
with respect to the \lcdm~prediction.
Even if the $\sigma_8$ values as determined by local measurements are an underestimate of the true value, nevertheless 
this can be a strong argument against MOD1.
Conversely, in MOD2, the DM abundance is fed by DE as the Universe evolves:
the non-linear evolution starts later and clustering is less prominent, making $\sigma_8$ smaller.

\looseness=-1
In Fig.~\ref{fig:sigma8_h0} we summarize the results on $H_0$ and $\sigma_8$ in the \lcdm\ , MOD1 and MOD2 models.
The left panel refers to CMB data only, while the right panel is for the ``ALL'' dataset.
The two bands show the intervals of the local determinations of $\sigma_8=0.75\pm0.03$ from Planck~\cite{Ade:2013lmv},
and $H_0=70.6\pm3.3$~\cite{Efstathiou:2013via}.
\looseness=-1
Both the plots show that MOD1 fails in obtaining high values of $H_0$ accompanied by low $\sigma_8$ values,
compatible with the local determinations of these parameters.
\looseness=-1
In MOD1 a larger $\sigma_8$ corresponds to a bigger DM/DE interaction rate, since the larger amount of DM in the early Universe
accelerates the evolution of the matter fluctuations at small scales.

\looseness=-1
MOD2 exhibits instead an opposite behavior, i.e.~lower values for $\sigma_8$ correspond to a stronger coupling in the dark sector
and possibly to high values of $H_0$, if $\wla$ is large.
In this sense, MOD2 appears to be preferred over MOD1, since in this context the cited tensions of $\sigma_8$ and $H_0$ can be solved.

\section{Conclusions}
\label{sec:concl}
\looseness=-1
In order to investigate the possibility of a non-gravitational interaction between DM and DE, we have performed cosmological tests
in a phenomenological model where DM can partially transfer its density to DE, or vice-versa~\cite{Murgia:2016ccp}.
The combination of a whole host of updated cosmological data allows to constrain
the evolution of the Universe at different redshifts and test the DM/DE interaction at different times.
If we consider the derived values of the Hubble parameter $H_0$ and of $\sigma_8$,
we find that MOD1
increases the tension with the low-redshift measurements of $H_0$ from HST~\cite{Efstathiou:2013via, Riess:2016jrr}
and, more significantly, with the SZ cluster counts and other local determinations of $\sigma_8$
\cite{Heymans:2012gg, Ade:2015fva, Ade:2013lmv}.
On the contrary, in MOD2, $\sigma_8$ is smaller than in the \lcdm~model, as a consequence of the DM to DE transfer, and cosmological determinations 
of $H_0$ and $\sigma_8$ are better reconciled with low-redshift probes.
A non-zero coupling between DE and DM, with an energy flow from the former to the latter, appears therefore to be in better 
agreement with cosmological data.

\section*{References}

\bibliographystyle{unsrt2}
\bibliography{cde2}

\end{document}